# NEW RESULTS ON SCRAMBLIMG USING THE MESH ARRAY

Sandhya Rangineni

**Abstract.** This paper presents new results on randomization using Kak's Mesh Array for matrix multiplication. These results include the periods of the longest cycles when the array is used for scrambling and the autocorrelation function of the binary sequence obtained from the cycles.

## INTRODUCTION

The mesh array of matrix multiplication was introduced by Kak in 1988 [1],[2]. It is able to multiply the matrices in only 2n-1 steps for two $n \times n$ matrices. In a new paper, this array has been proposed as a scrambling transformation [3]. Figure 1 presents the mesh array for multiplying two $4 \times 4$ matrices. Here we investigate some additional scrambling properties of the array and also consider triple matrix multiplication.

## PRELIMINARIES

When multiplying two matrices A and B (C=AB), the components of C are obtained in the following arrangement:

11  22  33  44

12  31  24  43

32  14  41  23

34  42  13  21

As shown in [3], the items of both standard array and mesh array will be written in an array as follows:

11  12  13  14  21  22  23  24  31  32  33  34  41  42  43  44

(                                                              )

11  22  33  44  12  31  24  43  32  14  41  23  34  42  13  21

By writing the above arrays into cycles, we can get period of the matrix of order 4. The period is nothing but the maximum of lengths of the cycles. The cycles of the matrix of order 4 are as follow:

= (11) (42) (12  22  31  32  14  44  21) (13  33  41  34  23  24  43)

Here the lengths of the cycles are {1,1,7,7}, and the period of the scrambling transformation = 7. We will now consider the longest cycle in each scrambling matrix. The period of the scrambling transformation will be the lcm of the cycles associated with the scrambling. Since the periods increase very rapidly, we shall consider only the longest cycles.

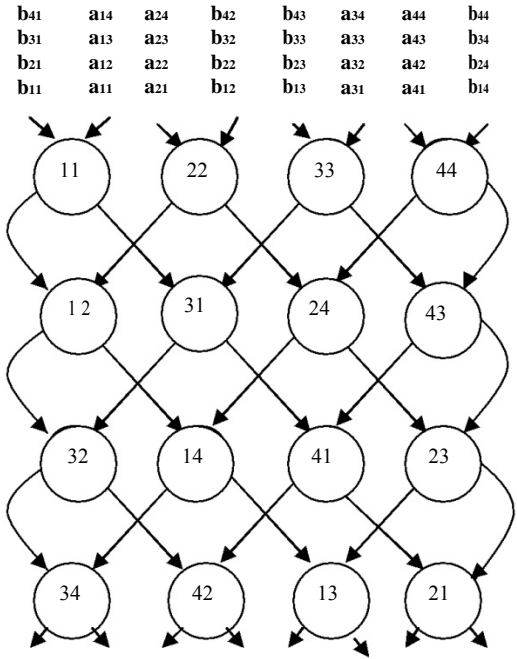

**Figure 1:** Mesh Architecture for multiplication of matrices A and B and store the result in C from [1]

We now consider further properties of the array for scrambling [4],[5], which has many applications in signal processing.

**Table 1:** Longest cycles for the matrices from order 2 to 100

| ORDER | LONGEST CYCLE |
|---|---|
| 2 | 3 |
| 3 | 7 |
| 4 | 7 |
| 5 | 20 |
| 6 | 23 |
| 7 | 19 |
| 8 | 27 |
| 9 | 79 |
| 10 | 31 |
| 11 | 88 |
| 12 | 46 |
| 13 | 150 |
| 14 | 180 |
| 15 | 103 |
| 16 | 197 |
| 17 | 242 |
| 18 | 270 |
| 19 | 121 |
| 20 | 220 |
| 21 | 438 |
| 22 | 402 |
| 23 | 367 |
| 24 | 455 |
| 25 | 478 |
| 26 | 362 |
| 27 | 667 |
| 28 | 514 |
| 29 | 262 |
| 30 | 678 |
| 31 | 697 |
| 32 | 414 |
| 33 | 507 |
| 34 | 620 |
| 35 | 512 |
| 36 | 492 |
| 37 | 1357 |
| 38 | 687 |
| 39 | 751 |
| 40 | 1110 |
| 41 | 1065 |
| 42 | 824 |
| 43 | 813 |
| 44 | 1221 |
| 45 | 912 |
| 46 | 1435 |
| 47 | 1347 |
| 48 | 877 |
| 49 | 2015 |
| 50 | 1391 |
| 51 | 1341 |
| 52 | 1090 |
| 53 | 2370 |
| 54 | 2182 |
| 55 | 974 |
| 56 | 2508 |
| 57 | 2064 |
| 58 | 2955 |
| 59 | 2146 |
| 60 | 2392 |
| 61 | 2452 |
| 62 | 2171 |
| 63 | 1448 |
| 64 | 2687 |
| 65 | 1957 |
| 66 | 4046 |
| 67 | 3069 |
| 68 | 1116 |
| 69 | 1501 |
| 70 | 3539 |
| 71 | 2219 |
| 72 | 2064 |
| 73 | 2542 |
| 74 | 3191 |
| 75 | 3194 |
| 76 | 5085 |
| 77 | 5329 |
| 78 | 2831 |
| 79 | 6060 |
| 80 | 3140 |
| 81 | 5390 |
| 82 | 3007 |
| 83 | 4786 |
| 84 | 6970 |
| 85 | 4012 |
| 86 | 3213 |
| 87 | 5143 |
| 88 | 7488 |
| 89 | 7685 |
| 90 | 5941 |
| 91 | 3383 |
| 92 | 6903 |
| 93 | 2521 |
| 94 | 4930 |
| 95 | 5869 |
| 96 | 6214 |
| 97 | 4419 |
| 98 | 3173 |
| 99 | 5150 |
| 100 | 7984 |

## Prime orders

The number of primes in the list of longest cycles has the following distribution:

001 – 100 ---- 16
101 – 200 ---- 15
201 – 300 ---- 10
301 – 400 ---- 11
401 – 500 ---- 5
501 – 600 ---- 3
601 – 700 ---- 8
701 – 800 ---- 5
801 – 900 ---- 4
901 – 1000---- 12

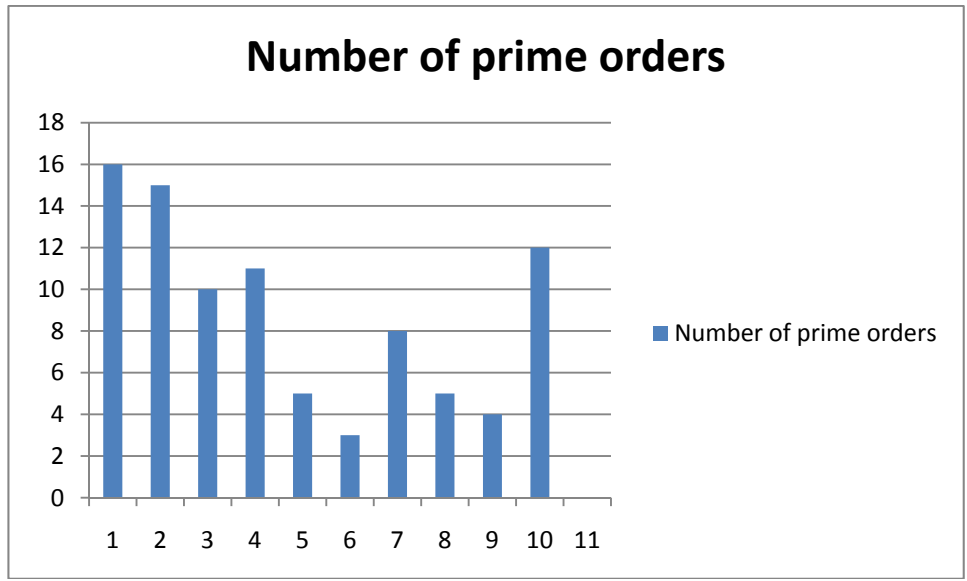

**Figure 2:** The graph for the number of prime orders from 0 to 1000

This in itself does not tell us how good are the randomness properties of the sequences of cycles associated with the mesh array. For this we will look at the autocorrelation function derived from the sequence.

## BINARY SEQUENCE FOR THE CYCLES

We can create a binary sequence of the longest cycles in terms of 1s and 0s where the even cycle is represented as 1 and odd cycle is represented as 0. The binary sequence for the cycles of orders 2 to 1000 is as follows:

11101101100001100000010000000000000000100000000100000000000000010000010000000000
00000000000000101000001100000000010001000000010101100000100000000000000000000000
10100000010000000000000000100000010000000100000110000000001000000001000010001000
00000000000000000000000001000100000000000000000000000100000010000000000010000
00001000000000001100000000000001000000000001000000000000000000000000000000000
00110111000000000000000000000101000000000000000000010000000000000000000100000
00001000000000000000000000000000000000000000001000000000000000000000000000000
00000000000000000000000000000000000000000010001000000000000000000001000001100100000
00000000000000000000000000010000000000000000000000011000010000000000000000000
00000000000010000000000000000000000000011000000000000000100000000000000000
00000000000000010000000000000000000000001000000100000000000100000000010000000
00000000000000000000000000000000000010000000100100000000001000000000000000
0000000000110000000000001000000101000000000100000000000000001000

## AUTOCORRELATION FUNCTION

Autocorrelation function is used to show the similarity between the observations as a function according to time. Here autocorrelation function is used to represent the variations of the cycles as a single function.

$$C(k) = 1/999 \sum A(i)*A(i+k) \text{ where } i=1 \text{ to } 999$$

Here, A(i) is the polar sequence of the cycles where 0 is converted as -1 in binary sequence and 1 remains same. The autocorrelation function for k ranging from 0 to 100 is shown in Figure 3.

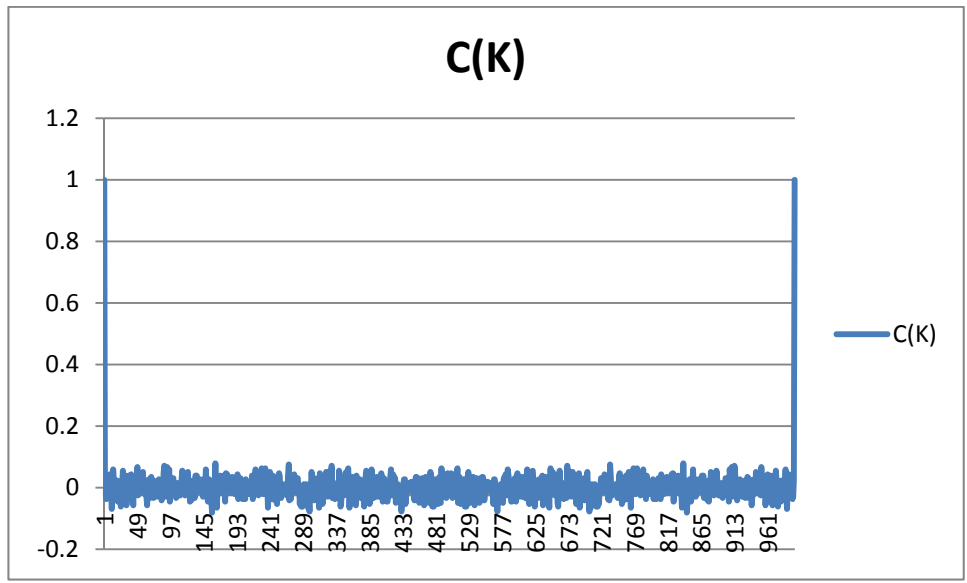

**Figure 3:** Autocorrelation function C(k) where k ranges from 0 to 1000

The autocorrelation function is effectively two valued which demonstrates that the sequence of orders is random.

# TRIPLE MATRIX MULTIPLICATION ON A MESH ARRAY

Now we consider the multiplication of three matrices of the same order in the manner of [6],[7]. Let A, X, B be the matrices to be multiplied and let us store the result in another matrix Y i.e., Y = A X B.

The computation of Y = A X B is decomposed into

- Z = XB
- Y = AZ

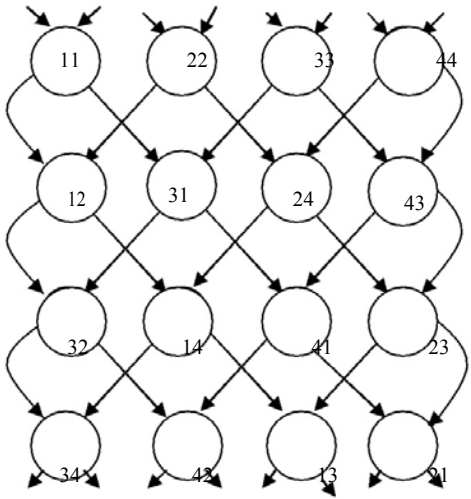

**Figure 4:** Mesh Architecture for Z = XB where 11 in the node represent $Z_{11}$

At time t=n the first row of the product XB is completed and the results of Z are stored in the nodes of first row and then we switch the X and B values to the second row transmitting downwards in the array. Now immediately after t=n the elements of the matrix, A follows the same path as X has passed. When t=2n, the product of XB in the second row and the product of AZ in the first row are done in parallel and the result of Z is stored in the nodes of second row and the result of AZ is stored in $Y_{ij}(1)$ respectively.

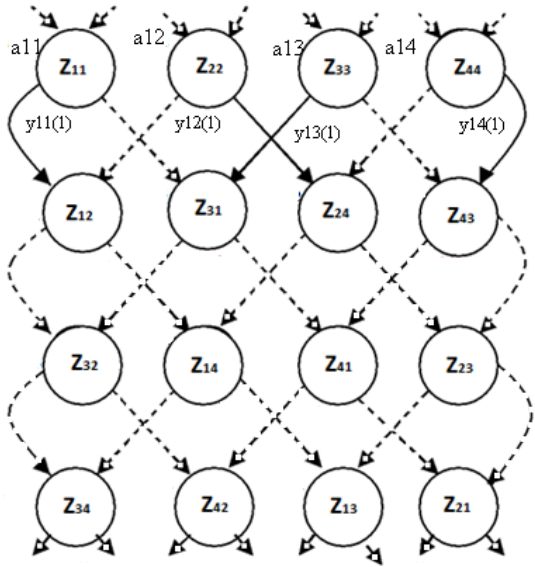

**Figure 5:** The triple matrix multiplication at time t=2n

Then at time t=2n, the results obtained are the Z values of second row and the Y values of first row. The process is continued till all the results are obtained.

$$Y_{ij} = Y_{ij}(1) + Y_{ij}(2) + Y_{ij}(3) + Y_{ij}(4)$$

At time t=4n, the mesh architecture for triple matrix multiplication is as follows:

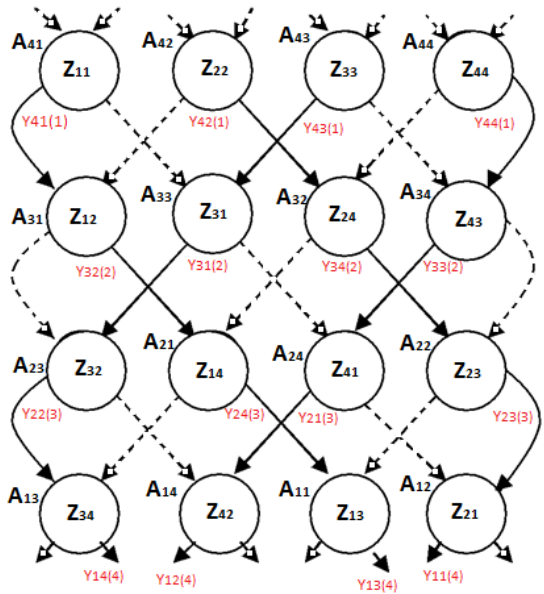

**Figure 6:** The mesh array representation for Y = AXB

# DISCUSSION

This article represents new results on scrambling using Kak's mesh array for matrix multiplication. These results includes the periods of the matrices multiplied, the binary sequence of the longest cycles for even and odd numbers, and the autocorrelation function for an even and odd sequence obtained from the order. The structure for triple matrix multiplication on a mesh array has also been presented.